# Cooling of trapped ions by resonant charge exchange

Sourav Dutta[*,†] and S. A. Rangwala

*Raman Research Institute, C. V. Raman Avenue, Sadashivanagar, Bangalore 560080, India*
(Submitted 22 May 2017, Revised 22 June 2017)

The two most widely used ion cooling methods are laser cooling and sympathetic cooling by elastic collisions (ECs). Here we demonstrate another method of cooling ions that is based on resonant charge exchange (RCE) between the trapped ion and the ultracold parent atom. Specifically, trapped $Cs^+$ ions are cooled by collisions with co-trapped, ultracold Cs atoms and, separately, by collisions with co-trapped, ultracold Rb atoms. We observe that the cooling of $Cs^+$ ions by Cs atoms is more efficient than cooling of $Cs^+$ ions by Rb atoms. This signals the presence of a cooling mechanism apart from the elastic ion-atom collision channel for the $Cs$–$Cs^+$ case, which is cooling by RCE. The efficiency of cooling by RCE is experimentally determined and the per-collision cooling is found to be two orders of magnitude higher than cooling by EC. The result provides the experimental basis for future studies on charge transport by electron hopping in atom-ion hybrid systems.

*Introduction.* — Cooling and trapping of dilute gases, both neutral and charged, have enabled extremely precise and controlled experimentation with these systems [1,2]. Simultaneous trapping and cooling of atoms and ions results in an ion-atom hybrid system that allows for exciting experimental possibilities. The hybrid system has lent itself to the studies of low energy ion-atom collisions [3–6], charge exchange reactions [3,5,7,8], sympathetic cooling of ions [4,9–11], cold chemical reactions [12], three body processes [13], non-destructive ion detection methods [14], vibrational state cooling of molecular ions [15] etc. and holds the promise for studies on charge transport [16,17], ion mobility [18], mesoscopic molecular ions [19], ion-atom photoassociation [20] and Feshbach resonance [21].

In these ion-atom hybrid systems, the ions are either laser cooled or sympathetically cooled by collisions with the ultracold atoms. Elastic collisions (ECs) are present in all ion-atom hybrid systems and therefore understanding it has attracted attention [4–6,9–11]. Recent experiments have studied the dependence of ion cooling on the size of the atomic ensemble and the atom-ion mass ratio [9–11]. Theoretical models [9,11,22–26] have also been developed to describe the cooling of trapped ions by ECs with cold atoms. In addition, another mechanism for cooling of ions by resonant charge exchange (RCE) has been proposed [9,17] but no experimental evidence of this cooling mechanism has yet been provided. In this article we experimentally show "swap cooling" of ions based on RCE between the trapped ion and the co-trapped, ultracold, parent atom. Such swap cooling is restricted to homonuclear systems, where it offers promising prospects. We determine the efficiency of cooling by RCE with respect to cooling by EC for trapped $Cs^+$ ions in ultracold Cs atoms and find the former to be higher. Our result confirms and quantifies experimentally the role of RCE in ion cooling by trapped parent atoms.

The demonstration of cooling by RCE is also the first step towards experimental realization of theoretical proposals [16,17] on the study of charge transport in the ultracold Na–$Na^+$ system and on studies of ion mobility [18]. The benefit of experiments with optically dark ions, e.g. $Na^+$, $Rb^+$, $Cs^+$ etc., is that complications in the ion-atom collision process due to the presence of ions in the excited state [7,12] can be avoided. However, dark ions cannot be laser cooled and therefore low energy collision experiments with such systems are scarce, although collisions in the keV range have been studied extensively [27,28]. It is only recently that low energy ($\lesssim 1$ eV) collisions were studied in the Rb–$Rb^+$ system [9] and the Na–$Na^+$ system [10], and sympathetic cooling of trapped dark ions by collisions with the parent neutral atoms was demonstrated. For such homonuclear systems, it was proposed in Ravi *et al.* [9] that the sympathetic ion cooling could be due to (*i*) ECs between the fast ion and the ultracold atoms or (*ii*) RCE between a fast ion and an ultracold atom, in which case the post collision ion is essentially at rest [see Fig. 1(a)], or a combination of both. The individual contribution of (*i*) and (*ii*) and therefore the role of RCE in the ion cooling process, essential for the physics proposed in Ref. [16], is yet to be shown experimentally.

To achieve this, we cool $^{133}Cs^+$ ions trapped in a Paul trap by collisions with ultracold $^{133}Cs$ atoms trapped in a magneto-optical trap (MOT) and, separately, by collisions with ultracold $^{85}Rb$ atoms trapped in a MOT. We observe that the cooling of $Cs^+$ ions by Cs atoms is more efficient than cooling of Cs$^+$ ions by Rb atoms. This cannot be explained by ion cooling models that consider only ECs between ions and atoms. The faster cooling in the Cs–Cs$^+$ case is attributed to RCE between Cs atoms and Cs$^+$ ions. We experimentally determine the efficiency of cooling by RCE and find the per-collision cooling to be two orders of magnitude higher than cooling by EC. This is remarkable given that the experiment is performed in a collision energy window where the ratio of elastic cross section ($\sigma^e$) to RCE cross section ($\sigma^x$) is close to its maximum [see Fig. 1(b)], favouring elastic collisions by a factor of ~ 70 [17,29,30]. The cooling of ions by RCE is not restricted to ions trapped in a Paul trap, where micromotion sets a limit to the lowest ion temperature that can be reached in a hybrid trap [6,31], and could possibly be extended to ions trapped in an optical dipole trap [32] where no such limitations exist.

---

* sourav@rri.res.in

† Present address: Department of Physics, Indian Institute of Science Education and Research (IISER) Bhopal, Bhopal – 462066, India. Email: sourav.dutta.mr@gmail.com

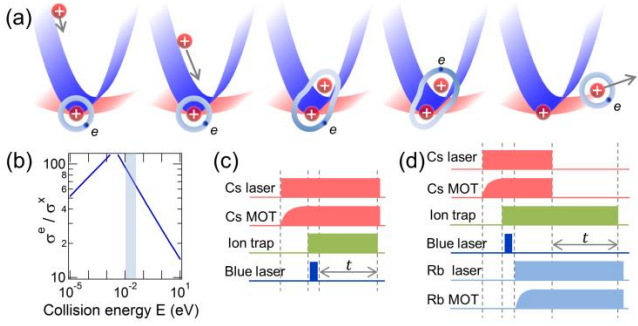

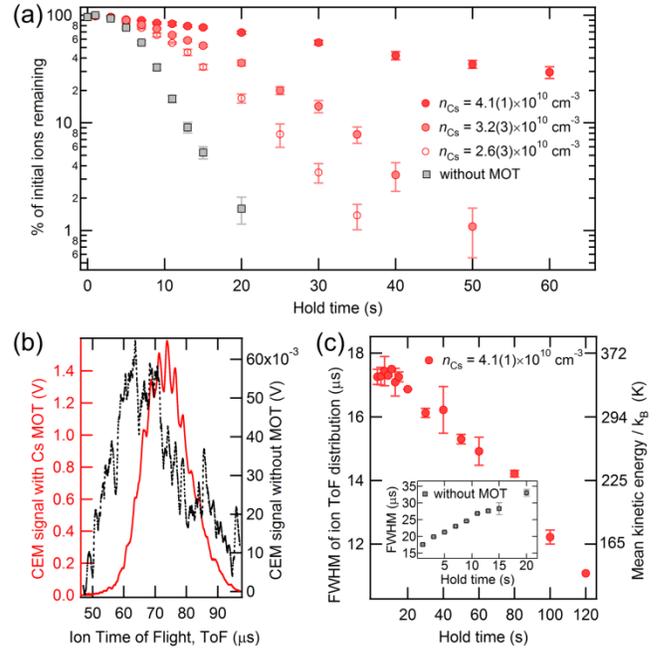

FIG. 1. (a) Pictorial representation of ion cooling by RCE. The Paul trap for ions is much deeper than the co-centered MOT for atoms. As the ion approaches the center of the Paul trap, its secular speed increases and it collides with an atom at rest. During the collision the outermost electron wavefunction is shared among both species, thus allowing RCE. Post collision, the fast moving atom is ejected from the MOT and a very cold ion occupies the center of the Paul trap. (b) The theoretically calculated [17,29,30] ratio of elastic to RCE cross section for the Cs–Cs$^+$ system. The behaviour of $\sigma^x$ changes from logarithmic in $E$ at high $E$ to $E^{-1/2}$ below a few meV [34], resulting in a sharp change in the trend of $\sigma^e/\sigma^x$. The shaded area represents the collision energy regime of the present experiments. (c) Timing sequence for the Cs–Cs$^+$ experiments. Results are presented in Fig. 2. (d) Timing sequence for the Rb–Cs$^+$ experiments. Results are presented in Figs. 3 & 4.

*Experimental set up.* — The apparatus [9,11,33] consists of a Paul trap for Cs$^+$ ions and MOTs for ultracold Cs and Rb atoms. Details of the MOTs and the Paul trap are provided in the accompanying Supplemental Material [34]. The Paul trap is well centered with the MOTs to ensure the most efficient cooling [9–11]. The $1/e^2$ radius (= 470 ± 25 μm) of the Cs and Rb MOTs are similar for all experiments. The MOT atom number (~ 5–10×10$^6$), and consequently the peak density ($n$), is tuned by changing the current through the atomic dispenser sources. To load the ion trap, a blue 473 nm laser is used to ionize Cs atoms in the $6P_{3/2}$ state that are already present in a Cs MOT. The ions are detected using a channel electron multiplier (CEM) and the time-of-flight (ToF) to the detector is used to differentiate ions of different species.

FIG. 2. (a) Decay of Cs$^+$ ions from the ion trap when held with (circles) and without (squares) a Cs MOT. The lifetime of Cs$^+$ ions in the ion trap increases as the Cs MOT density $n_{Cs}$ increases. The increase in lifetime is due to cooling of trapped Cs$^+$ ions by Cs atoms in the MOT. (b) The CEM signal when ions are extracted after 15 s of hold time either in the presence of a Cs MOT (solid line, left axis) or in the absence of a MOT (dotted line, right axis). The width of the ion ToF distribution is lower in presence of the Cs MOT. The peak position shifts to higher ToF for colder ions due to paraxial lensing. (c) In presence of a Cs MOT, the FWHM of the ion ToF distribution (obtained by fitting to a Gaussian function) decreases as the hold time increases suggesting cooling of ions by the Cs atoms. The ions' mean kinetic energy (in temperature units) is plotted on the right axis. Inset: In absence of a Cs MOT, the FWHM of the ion ToF distribution increases as the hold time increases signifying heating of ions.

*Cooling of Cs$^+$ ions by ultracold Cs atoms.* — The experimental sequence is depicted in Fig. 1(c). First the shutter in the path of the Cs laser beams is opened and a Cs MOT is allowed to load till saturation, the rf and dc ion trap voltages are turned on and Cs$^+$ ions are created by turning on the blue laser briefly which loads ~ 1000 Cs$^+$ ions in the ion trap. Subsequently, either the shutter is kept open allowing the Cs$^+$ ions to interact with the Cs MOT atoms or the shutter is closed, emptying the Cs MOT. The ions are held in the ion trap for a predetermined hold time $t$ and then the surviving ions are extracted from the ion trap and detected using the CEM. The hold time is changed and the sequence is repeated. The experiment is then repeated for different values of Cs MOT density ($n_{Cs}$).

The result of the experiments is shown in Fig. 2. Note that for Figs. 2-4, standard deviation (1σ) error bars are shown and are smaller than the data points where not visible. From Fig. 2(a), it is clear that the lifetime of the Cs$^+$ ions in the ion trap increases when a Cs MOT is present. This is due to the cooling (i.e. reduction in secular speed) of Cs$^+$ ions by collisions with Cs MOT atoms. Here, secular speed refers to the speed of an ion oscillating at its secular frequency. The cooling of ions occurs because the atoms are localized in a region much smaller than the volume occupied by the ions and are placed precisely at the center of the ion trap – a geometry that always results in reduction of the secular speed of a trapped ion after collision thereby cooling the ion [9,11]. Figure 2(a) also shows that the cooling effect increases as the Cs MOT density $n_{Cs}$ increases. This is a result of increase in the ion-atom collision rate as $n_{Cs}$ increases, leading to more efficient ion cooling.

In Fig. 2(b) we show an example of the ion ToF distribution, at hold time $t$ = 15 s, for the "with MOT" and "without MOT" cases – the full width at half maximum (FWHM) of the distribution reduces when the Cs MOT is present, providing independent evidence of ion cooling. The FWHM reduces because the extent of the secular orbit of a trapped ion is reduced due to ion cooling by the MOT atoms resulting in compression in the phase-space. Figure 2(c) shows that the FWHM of the ions' ToF distribution keeps reducing as the hold time increases – a result of systematic reduction in the ions' mean kinetic energy due to cooling by the MOT atoms. See Supplemental Materials [34] for details regarding the determination of ion temperature. Note that in absence of the MOT the FWHM increases with increasing hold time [inset of Fig. 2(c)] suggesting heating of ions due, primarily, to trap imperfections and, to a much lower extent, to collisions with the background gas.

*Rationale for the Rb–Cs$^+$ experiment.* — While the above results confirm Cs$^+$ ion cooling by Cs MOT atoms, there is no experimental signature that can distinguish between the EC and the RCE channels for cooling. For the discriminatory test we cool trapped Cs$^+$ ions with ultracold Rb atoms. Since Cs$^+$ and Rb are different species, RCE between Cs$^+$ and Rb is not possible. Moreover, related calculations [35], related experiments [11] and the results below show that rate of non-resonant charge exchange (nRCE) at the low (≤ 1 eV) collision energy is negligible. Therefore, for the cooling of Cs$^+$ ions by Rb MOT, the ion cooling is through ECs only, which is explained accurately by theoretical models [25,26]. The EC cooling rates ($k^e_{Cs}$, $k^e_{Rb}$) in these models depend on the atom-ion mass ratios (ξ = 1.00, ξ = 0.64) and the EC cross sections ($\sigma^e_{Cs}$, $\sigma^e_{Rb}$) [29] for the two cases (Cs–Cs$^+$, Rb–Cs$^+$). The effective ratio of EC cross section is $(\sigma^e_{Cs}/\sigma^e_{Rb})_{eff}$ = 1.23 (see Supplemental Material [34]). For our MOTs, which are localized but finite in size, the model [25,26] predicts 0.62 ≲ $(k^e_{Cs}/k^e_{Rb})$ ≲ 1.28 (see [34]). Thus, the theoretical (th) upper bound (ub) is $(k_{Cs}/k_{Rb})^{ub}_{th}$ = 1.28 – an experimental value higher than this would bring out the role of RCE. Here $k_{Cs}$ and $k_{Rb}$ are the total (EC + RCE) cooling rates for the Cs–Cs$^+$ and Rb–Cs$^+$ cases, respectively.

*Cooling of Cs$^+$ ions by ultracold Rb atoms.* — The experimental sequence is depicted in Fig. 1(d) and is similar to the Cs–Cs$^+$ experiments but with one preparation difference. After the loading of the Cs$^+$ ions, the Cs MOT is kept operative and the Rb laser beams are allowed in and the Rb MOT is allowed to load till saturation. During this time, which is 10 s, the Cs$^+$ ions interact with both the Cs and the Rb MOTs. Instead, if the Cs MOT is emptied immediately after the Cs$^+$ ions are loaded, there is a severe loss of Cs$^+$ ions [see the "without MOT" case in Fig. 2(a)], while the Rb MOT loads. After the 10 s duration (this is taken to be $t$ = 0 s), the Cs MOT (Rb MOT) is emptied by blocking the Cs (Rb) laser beams and the Cs$^+$ ions are held in the presence of the Rb MOT (Cs MOT) for a variable hold time $t$ to get the ion decay curve with Rb MOT (with Cs MOT). Both MOTs are emptied at $t$ = 0 s to get the "without MOT" data. The surviving ions are detected using the CEM and the number of ions and the FWHM of the ion ToF distribution determined. Notably, the detected ion ToF distribution is consistent with Cs$^+$ ions and not consistent with the ToF profile of Rb$^+$ ions (on average, Rb$^+$ ions would arrive 14 μs before the Cs$^+$ ions). No confirmed detection of Rb$^+$ ions could be made even when the ion trap parameters were adjusted to trap Rb$^+$ ions efficiently – we therefore conclude that all detected ions are Cs$^+$ and that nRCE (Cs$^+$ + Rb → Cs + Rb$^+$) rates are very small.

In Fig. 3 we plot the number of Cs$^+$ ions detected at different hold times when the ions are held in presence and absence of the Rb or Cs MOT. The density of Rb MOT [$n_{Rb}$ = 4.0(1)×10$^{10}$ cm$^{-3}$] is similar to density of the Cs MOT [$n_{Cs}$ = 3.8(2)×10$^{10}$ cm$^{-3}$]. Compared to the "without MOT" case, a larger number of Cs$^+$ ions survive when the Rb MOT is present – this is due to cooling by ECs with ultracold Rb atoms. The experiment with Cs$^+$ ions in presence of a Cs MOT results in an even longer ion lifetime. On fitting the $t$ ≥ 8 s data to an expression of the form $Ae^{-\gamma t}$, we get $\gamma_0$ = 0.34(1) s$^{-1}$, $\gamma_{Rb}$ = 0.17(1) s$^{-1}$ and $\gamma_{Cs}$ = 0.047(3) s$^{-1}$ for the "without MOT", the "with Rb MOT" and the "with Cs MOT" cases, respectively. $\gamma_0$, $\gamma_{Rb}$ and $\gamma_{Cs}$ depend on the ion heating rates in the respective

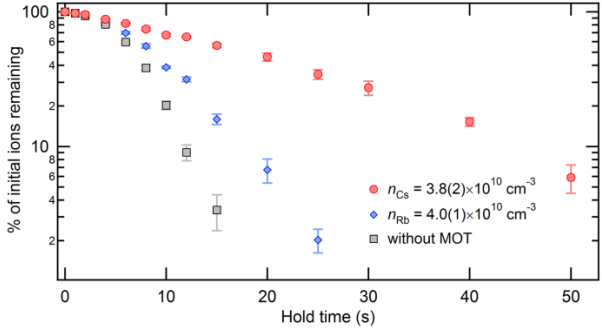

FIG. 3. Decay of $Cs^+$ ions from the ion trap when held with a Rb MOT (rhombuses), a Cs MOT (circles) or without any MOT (squares). The lifetime $Cs^+$ in the ion trap increases in presence of either of the two MOTs but the enhancement in lifetime is higher when the Cs MOT is present. The densities of the Cs and Rb MOTs are similar.

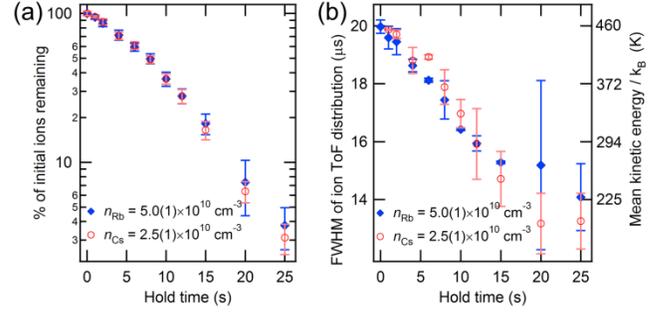

FIG. 4. (a) Decay of $Cs^+$ ions from the ion trap when held in presence of a Rb MOT (rhombuses) of density $n_{Rb} = 5.0(1) \times 10^{10}$ $cm^{-3}$ or in presence of a Cs MOT (circles) of density $n_{Cs} = 2.5(1) \times 10^{10}$ $cm^{-3}$. The decay curves are essentially identical. (b) The FWHM of the ion ToF distribution, and the corresponding mean kinetic energy plotted on the right axis, when held in presence of a Rb MOT (rhombuses) or in presence of a Cs MOT (circles). The FWHM of the ion ToF distribution decreases as the hold time increases suggesting cooling of ions in both cases and at similar rates.

cases – therefore $-(\gamma_{Cs} - \gamma_0)$ and $-(\gamma_{Rb} - \gamma_0)$ represent the cooling due to Cs MOT and Rb MOT, respectively. The ratio $(\gamma_{Cs} - \gamma_0)/(\gamma_{Rb} - \gamma_0) = 1.72(2)$ provides an experimental estimate of $(k_{Cs}/k_{Rb})$, and is higher than the theoretical upper bound $(k_{Cs}/k_{Rb})_{th}^{ub} = 1.28$ based on ECs only. We therefore conclude that an additional cooling channel is active in the $Cs-Cs^+$ case and it must be RCE since it can bring a fast $Cs^+$ ion to rest in a single ion-atom collision [see Fig. 1(a)].

This cooling beyond the bounds of EC cooling must be attributed to a new process. There are only two mechanisms, other than RCE, which need to be considered. The first mechanism is one in which a fast ion collides with an atom at rest and kinetically excites the atom to an excited electronic state. Such a collision could result in ion cooling. However, such collisions are energetically forbidden in our experiment since the mean kinetic energy of a trapped ion is < 0.05 eV whereas the minimum energy required for an electronic excitation ($6S_{1/2} \rightarrow 6P_{1/2}$) of Cs is ~ 1.39 eV. The other mechanism is one in which a fast ion collides with an atom at rest and transfers its kinetic energy resulting in a hyperfine excitation in the atom. A single atomic hyperfine changing collision can change the energy of the ion by only ~ 38 μeV (≡ 9.2 GHz atomic hyperfine splitting). However, if active, in our experiments the atomic hyperfine state changing collisions would be a heating mechanism for the ions. This is because the atoms in a MOT are in the upper ($6S_{1/2}$ F = 4) hyperfine state and therefore would transfer energy to the ion while making a transition to the lower ($6S_{1/2}$ F = 3) hyperfine state. The electronic or hyperfine state changing collisions therefore cannot explain the observed cooling. Further, the results of the experiment are in good agreement with expectations of cooling due to the RCE process, allowing us to attribute the faster cooling in $Cs-Cs^+$ case to the RCE process.

*Quantifying the RCE cooling rate.* — The above determination of $(k_{Cs}/k_{Rb})$ based on difference in ion lifetime depends on the fitted equation. To avoid this model dependence, we develop an alternative experimental strategy to quantify the role of RCE. We tune $n_{Cs}$ and $n_{Rb}$ to determine a pair of values for which the ion decay curves for both the $Cs-Cs^+$ and $Rb-Cs^+$ cases are essentially identical [Fig. 4(a)]. This is observed to occur when the ratio $n_{Rb}/n_{Cs} = 2.0(2)$. Further it is seen that at each hold time the FWHM of the ToF distributions for the two cases also coincide [Fig. 4(b)], suggesting matched cooling rates under these conditions. Since $k_{Rb} \propto n_{Rb}$ and $k_{Cs} \propto n_{Cs}$, these figures suggest that $(k_{Cs}/k_{Rb}) = 2.0(2)$. This value is close to 1.72(2) obtained earlier and is higher than 1.28 confirming that cooling by RCE is active in the $Cs-Cs^+$ case. Thus $k_{Cs}$ has contributions from both ECs ($k_{Cs}^e$) and RCE ($k_{Cs}^x$), i.e. $k_{Cs} = k_{Cs}^e + k_{Cs}^x$, while $k_{Rb}$ has contribution only from ECs, i.e. $k_{Rb} = k_{Rb}^e$. So, we have $(k_{Cs}^e + k_{Cs}^x) = 2 k_{Rb}^e$. On using $0.62 \lesssim (k_{Cs}^e/k_{Rb}^e) \lesssim 1.28$ (see [34]), we get $0.56 \lesssim (k_{Cs}^x/k_{Cs}^e) \lesssim 2.23$, i.e. the effective cooling rates for cooling by RCE and cooling by ECs are of similar magnitude. However, as discussed below, the cooling efficiency for a single collision is much greater in the case of RCE cooling.

The cooling rate $k_{Cs}$ can be defined as $k_{Cs} \equiv k_{Cs}^e + k_{Cs}^x \approx \delta E_{Cs}^e n_{Cs} \sigma_{Cs}^e v + \delta E_{Cs}^x n_{Cs} \sigma_{Cs}^x v$, where $\delta E_{Cs}^x$ ($\delta E_{Cs}^e$) is the energy lost by a $Cs^+$ ion in a single RCE (EC) collision with a Cs atom (see [34]), $\sigma_{Cs}^x$ ($\sigma_{Cs}^e$) is the RCE (EC) cross section and $v$ is the speed of the ion before collision. Taking a ratio between $k_{Cs}^x$ and $k_{Cs}^e$, we get $\delta E_{Cs}^x/\delta E_{Cs}^e = (\sigma_{Cs}^e/\sigma_{Cs}^x)(k_{Cs}^x/k_{Cs}^e)$. The theoretically calculated value of ($\sigma_{Cs}^e/\sigma_{Cs}^x$) varies non-monotonically with collision energy [17,29,30] with a maxima (~ 100) around 10 meV collision energy [see Fig. 1(b)]. Our experiments are conducted near this region at collision energies of ~ 30 meV ($\equiv$ 350 K) where ($\sigma_{Cs}^e/\sigma_{Cs}^x$) ~ 69 [17,29,30]. On using 0.56 $\lesssim$ ($k_{Cs}^x/k_{Cs}^e$) $\lesssim$ 2.23 obtained above, we get 39 $\lesssim$ $\delta E_{Cs}^x/\delta E_{Cs}^e$ $\lesssim$ 154. The result shows that the cooling per RCE collision is dramatically more efficient than cooling per EC. This is consistent with the hopping of an electron from the parent atom to the daughter ion in a single collision and results in rapid cooling despite the lower RCE cross section.

*Comments from supplementary experiments.* — We conducted control experiments with Cs–Rb$^+$ and Rb–Rb$^+$ and found that cooling in the Rb–Rb$^+$ is more efficient. While this is suggestive of cooling by RCE in the Rb–Rb$^+$ case, the data presented in Figs. 3 and 4 are better demonstration of RCE. This is because even when considering only ECs, one expects that cooling of Rb$^+$ by Rb will be more efficient than cooling of Rb$^+$ by Cs, since cooling by ECs is more effective when the mass of the neutral atom is lower [25,26]. We have also conducted experiments with Rb–Cs$^+$ and Rb–Rb$^+$ and found cooling to be more efficient in the latter, suggesting the presence of the additional RCE cooling mechanism in the homonuclear Rb–Rb$^+$ case.

*Conclusion.* — In conclusion, we demonstrate a very efficient method of ion cooling based on RCE. We show cooling of Cs$^+$ ions, trapped in a Paul trap, by collisions with localized, precisely centered, ultracold Cs and Rb atoms trapped in their respective MOTs. The cooling of Cs$^+$ ions by Cs atoms is more efficient than cooling by Rb atoms, the reason for which is RCE in the former case – similar cooling should occur in all other parent-daughter systems. The experimentally determined per-collision cooling in case of RCE is found to be two orders of magnitude higher than cooling by EC. This has direct implications for reaching the ultracold temperature regime, i.e. 100 μK or lower, in ion-atom collisions. For example, in absence of ion heating mechanisms, an ion at 350 K would require ~ 240 ECs to cool down to 100 μK whereas the same effect can result from a single RCE collision. Our findings also suggest that an ion can be co-trapped with atoms for very long durations if a sufficiently dense and localized gas of ultracold atoms is available, thereby allowing controlled studies of ion-atom collisions. The result also establishes the experimental basis for future experiments on charge mobility [16], mesoscopic molecular ions [19] and impurity physics in many body systems.

*Acknowledgments.* — We are happy to acknowledge helpful discussions with Daniel Comparat, Olivier Dulieu and Anders Kastberg. S.A.R. acknowledges support from the Indo-French Centre for the promotion of Advanced Research-CEFIPRA, project 5404-1. S.D. acknowledges support from the Department of Science and Technology (DST), India in the form of the DST-INSPIRE Faculty Award (IFA14-PH-114).

**Supplemental Material**

*Details of the Cs and Rb MOTs.* — Each MOT is formed by three pairs of mutually orthogonal retro-reflected laser beams that intersect at the center of the Paul trap. The laser beams contain light at cooling and repumping frequencies and the magnetic field gradient is ~ 16 Gauss/cm. The cooling beam for Cs (Rb) is detuned from the $6S_{1/2}$ $F = 4 \leftrightarrow 6P_{3/2}$ $F' = 5$ ($5S_{1/2}$ $F = 3 \leftrightarrow 5P_{3/2}$ $F' = 4$) transition by -16.5 MHz (-13 MHz) while the repumping beam is on resonance with the $6S_{1/2}$ $F = 3 \leftrightarrow 6P_{3/2}$ $F' = 4$ ($5S_{1/2}$ $F = 2 \leftrightarrow 5P_{3/2}$ $F' = 3$) transition. The Cs and Rb MOTs are very well overlapped spatially and are located at the center of the Paul trap. The loading of the Cs and Rb MOTs are independently controlled by mechanical shutters placed in the respective beam paths. The MOT atom number and size is determined from the MOT fluorescence recorded on a calibrated photomultiplier tube and a CCD camera, respectively. The statistical (i.e. random) error in the measurement of the MOT densities is typically below 10% and is quoted in the manuscript.

Note on systematic errors in the measurement of MOT densities: Since we image and measure the fluorescence of the Rb and Cs MOTs using the same optical set up, many sources of systematic errors (such as solid angle subtended, losses in light collection and focussing optics, etc.) are common to both the Rb and Cs MOTs – therefore, these errors do not alter the relative density measurement. The other factors which come into the determination of atomic density are the detuning of the cooling lasers and their optical powers. The detuning is determined to accuracy better than 1 MHz and the optical powers are also measured with uncertainty less than 3%. In the worst case scenario, these factors together lead to an uncertainty (systematic) of around 12% in the estimation of relative density.

*Details of the Paul trap.* — The linear Paul trap is operated by applying a sinusoidal radiofrequency (rf) voltage of amplitude 95 V and frequency 400 kHz to the four rod electrodes such that the voltage on adjacent electrodes are 180° out of phase, while the ring electrodes that serve as end cap are biased at 80 V. This results in a trap with radial (axial) secular frequency of 120 kHz (21 kHz) and a trap depth of ~ 1 eV. At these settings the Cs$^+$ ions are trapped efficiently but trapping of Rb$^+$ ions is inefficient. We initially load ~ 1000 Cs$^+$ ions in the ion trap

– this avoids saturation of the ion detector and keeps the ion density low enough such that ion-ion interaction is negligible. The extent of the trapped ions is much larger than the size of the MOTs. To detect the ions, the voltage on one of the end cap electrodes is suddenly changed from 80 V to -3 V which launches the ions along the axis of the time-of-flight (ToF) mass spectrometer at the end of which is a channel electron multiplier (CEM) where ions are detected destructively. The measured ion signal is proportional to the number of trapped ions just before extraction. Ions of different species can be differentiated based on their arrival time at the detector. The mean ToF of $Cs^+$ ions is 14 μs higher than $Rb^+$ ions ($Rb^+$ ions were created in a separate experiment by ionizing Rb MOT atoms). The half width at half maxima (HWHM) of the ToF distribution for $Cs^+$ ions is ≤ 10 μs and that of $Rb^+$ ions is ≤ 7 μs, where the upper bound in the HWHM occurs when ions are not cooled.

*Estimate of mean kinetic energy ($\equiv k_B T$) of the ions.* — The depth of the ion trap is ~ 1 eV, which sets the maximum speed $v_{max}$ (= 1200 m/s) that a $Cs^+$ ion can have and still remain trapped. Assuming a Maxwell Boltzmann (MB) distribution for the speed of the ions and requiring that more than 99.9% of the ions have speeds less than $v_{max}$, we find that the constraint is satisfied by a MB distribution at temperature 1250 K and most probable speed $v_p$ = 395 m/s. Thus, the maximum possible temperature of trapped $Cs^+$ ions is 1250 K. The ions at 1250 K would generate a ToF distribution of width 33 μs – the maximum width seen experimentally [inset Fig. 2C]. It can be easily shown that the width ($w$) of the ToF distribution depends on the square root of the ion's kinetic energy and therefore on the square root of temperature ($T$) i.e. $w \approx \alpha\sqrt{T}$. This proportionally has also been tested using molecular dynamics simulations [9]. The proportionally constant $\alpha$ (= $0.93 \times 10^{-6}$ s $K^{-1/2}$) is derived from the constraint that $w$ = 33 μs when $T$ = 1250 K. The experimentally measured ToF width $w$ can therefore be converted to the ion temperature using the relation $T \approx w^2/\alpha^2$. This method of determining temperature works for $T \gtrsim 20$ K when the ions are relatively dilute. For lower temperatures the ion-ion interaction becomes non-negligible and the simple relation no longer holds.

*Determination of effective EC cross section.* — The ion-atom elastic collision (EC) cross section is given by [17]: $\sigma^e = \pi(1+\pi^2/16)\,(\mu C_4^2/\hbar^2)^{1/3}\, E^{-1/3}$ [see Fig. 1S]. Here μ is the ion-atom reduced mass, $C_4$ is the coefficient of the $-C_4/2R^4$ ion-atom interaction potential and depends linearly on the polarizability $\alpha$ of the neutral atom, and $E$ is the (kinetic) collision energy. Considering the ground state polarizability [29] of Rb (Cs) to be 319 a.u. (400 a.u.) and the fact that our Rb (Cs) MOT contains ~ 12% (~ 5%) atoms in the $5P_{3/2}$ ($6P_{3/2}$) state with polarizability 870 a.u. (1648 a.u.); we determine that the effective polarizability in Cs–$Cs^+$ case is 1.20× higher than the Rb–$Cs^+$ case. The value of μ in the Cs–$Cs^+$ case is 1.28× higher than the Rb–$Cs^+$ case and $E$ (which is determined by the ion energy only because the atoms are essentially at rest) is the same in both cases. These lead to an effective ratio of EC cross section between the Cs–$Cs^+$ case and the Rb–$Cs^+$ case ($\sigma_{Cs}^e/\sigma_{Rb}^e)_{eff}$ = 1.23, for our experimental parameters.

*Theoretically calculated RCE cross section.* — In contrast to the EC cross $\sigma^e$ which varies as $E^{-1/3}$ for all energies greater than a μeV [see Fig. 1S], the resonant charge exchange (RCE) cross section $\sigma^x$ varies as $(-a\ln E + b)$ at high collision energies and as $E^{-1/2}$ at low collision energies [17], where $a$ and $b$ are constants. We use ref. [30] to calculate the values $\sigma^x$ for $E > 2.7$ meV ($\equiv 10^{-4}$ Hartree) and the method of ref. [17], i.e. $\sigma^x \sim \sigma_L/4$, for $E < 2.7$ meV. Here $\sigma_L = \pi\sqrt{2C_4}E^{-1/2}$ is the Langevin cross section. Note that $\sigma^e \gg \sigma^x$ across the entire energy range. The change in behaviour of $\sigma^x$ around 2.7 meV results in a sharp change in the behaviour of $\sigma^e/\sigma^x$ around 2.7 meV as shown in Fig. 1(b) of the manuscript.

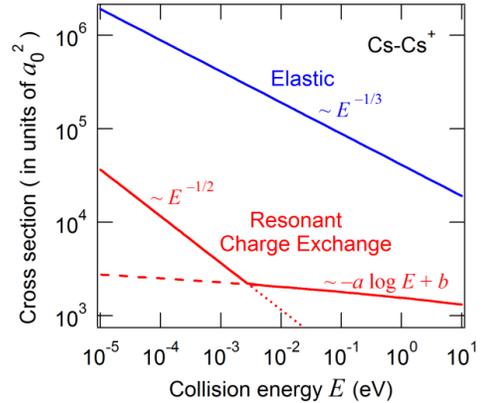

FIG. 1S. Calculated values of EC cross section (blue line) [17] and RCE cross section (solid red line) [17,30] as a function of collision energy.

*Theoretical estimate of EC cooling rate.* — Consider cooling of a single ion from some initial energy $E_i$ to a final energy $E_f$ via ECs with cold atoms. This would require $N$ ion-atom collisions, each collision taking away a fraction of the energy. To determine the gross effect over many collisions, we define $\delta E^e$ as the effective energy lost per EC such that $\delta E^e$ times the number of collisions gives the total energy lost. The effective EC cooling rate $k^e$ is obtained by multiplying $\delta E^e$ with the ion-atom collision rate i.e. $k^e \approx \delta E^e\, n\sigma^e v$, where $n$ is atomic number density, $\sigma^e$ is the ion-atom EC cross-section and $v$ is the speed of the ion. We will denote $\delta E_{Cs}^e$ ($\delta E_{Rb}^e$) as the effective energy lost by a Cs$^+$ ion per EC collision with a Cs (Rb) atom. As explained in Refs. [25,26], $\delta E_{Cs}^e$ and $\delta E_{Rb}^e$ depend on the respective atom-ion mass ratios $\xi$ ($\xi$ = 1.00 in the Cs–Cs$^+$ case, $\xi$ = 0.64 in the Rb–Cs$^+$ case). For a uniform distribution of cold buffer gas atoms the model predicts $\delta E_{Cs}^e \sim 0.5\, \delta E_{Rb}^e$ for a Paul trap (see Fig. 8 of Ref. [26]). On the other hand, if the atoms are localized in an infinitesimally small volume, the ion-atom collisions start mimicking collisions in free space (since micromotion is negligible) and the model predicts $\delta E_{Cs}^e \sim 1.04\, \delta E_{Rb}^e$ (see Fig. 8 of Ref. [26]). Therefore for our MOTs, which are localized but finite in size, the model would predict $0.5 \lesssim (\delta E_{Cs}^e / \delta E_{Rb}^e) \lesssim 1.04$. Further, the EC cooling rate $k_{Rb}^e$ in the Rb–Cs$^+$ case can be defined as $k_{Rb}^e \approx \delta E_{Rb}^e\, n_{Rb}\, \sigma_{Rb}^e\, v$, where $v$ is the speed of the ion before collision. Similarly, $k_{Cs}^e \approx \delta E_{Cs}^e\, n_{Cs}\, \sigma_{Cs}^e\, v$. For the same atomic densities (i.e. $n_{Cs} = n_{Rb}$), the ratio of EC cooling rate $(k_{Cs}^e / k_{Rb}^e) = (\sigma_{Cs}^e / \sigma_{Rb}^e)(\delta E_{Cs}^e / \delta E_{Rb}^e)$. Using $(\sigma_{Cs}^e / \sigma_{Rb}^e)_{eff} = 1.23$ and $0.5 \lesssim (\delta E_{Cs}^e / \delta E_{Rb}^e) \lesssim 1.04$, we get $0.62 \lesssim (k_{Cs}^e / k_{Rb}^e) \lesssim 1.28$ as the theoretical bounds for the ratio of EC cooling rates.

**References**

[1] A. D. Ludlow, M. M. Boyd, J. Ye, E. Peik, and P. O. Schmidt, Rev. Mod. Phys. **87**, 637 (2015).
[2] W. M. Itano, J. C. Bergquist, J. J. Bollinger, and D. J. Wineland, Phys. Scr. **T59**, 106 (1995).
[3] A. T. Grier, M. Cetina, F. Oručević, and V. Vuletić, Phys. Rev. Lett. **102**, 223201 (2009).
[4] C. Zipkes, S. Palzer, C. Sias, and M. Köhl, Nature **464**, 388 (2010).
[5] S. Schmid, A. Härter, and J. Hecker Denschlag, Phys. Rev. Lett. **105**, 133202 (2010).
[6] Z. Meir, T. Sikorsky, R. Ben-shlomi, N. Akerman, Y. Dallal, and R. Ozeri, Phys. Rev. Lett. **117**, 243401 (2016).
[7] L. Ratschbacher, C. Zipkes, C. Sias, and M. Köhl, Nat. Phys. **8**, 649 (2012).
[8] S. Haze, R. Saito, M. Fujinaga, and T. Mukaiyama, Phys. Rev. A **91**, 032709 (2015).
[9] K. Ravi, S. Lee, A. Sharma, G. Werth, and S. A. Rangwala, Nat. Commun. **3**, 1126 (2012).
[10] I. Sivarajah, D. S. Goodman, J. E. Wells, F. A. Narducci, and W. W. Smith, Phys. Rev. A **86**, 063419 (2012).
[11] S. Dutta, R. Sawant, and S. A. Rangwala, Phys. Rev. Lett. **118**, 113401 (2017).
[12] F. H. J. Hall, M. Aymar, N. Bouloufa-Maafa, O. Dulieu, and S. Willitsch, Phys. Rev. Lett. **107**, 243202 (2011).
[13] A. Krükow, A. Mohammadi, A. Härter, J. Hecker Denschlag, J. Pérez-Ríos, and C. H. Greene, Phys. Rev. Lett. **116**, 193201 (2016).
[14] S. Dutta and S. A. Rangwala, Phys. Rev. A **94**, 053841 (2016).
[15] W. G. Rellergert, S. T. Sullivan, S. J. Schowalter, S. Kotochigova, K. Chen, and E. R. Hudson, Nature **495**, 490 (2013).
[16] R. Côté, Phys. Rev. Lett. **85**, 5316 (2000).
[17] R. Côté and A. Dalgarno, Phys. Rev. A **62**, 012709 (2000).
[18] G. Heiche and E. A. Mason, J. Chem. Phys. **53**, 4687 (1970).
[19] R. Côté, V. Kharchenko, and M. D. Lukin, Phys. Rev. Lett. **89**, 093001 (2002).
[20] A. Rakshit and B. Deb, Phys. Rev. A **83**, 022703 (2011).
[21] Z. Idziaszek, T. Calarco, P. S. Julienne, and A. Simoni, Phys. Rev. A **79**, 010702(R) (2009).
[22] R. G. De Voe, Phys. Rev. Lett. **102**, 063001 (2009).
[23] C. Zipkes, L. Ratschbacher, C. Sias, and M. Köhl, New J. Phys. **13**, 053020 (2011).
[24] K. Chen, S. T. Sullivan, and E. R. Hudson, Phys. Rev. Lett. **112**, 143009 (2014).
[25] B. Höltkemeier, P. Weckesser, H. López-Carrera, and M. Weidemüller, Phys. Rev. Lett. **116**, 233003 (2016).
[26] B. Höltkemeier, P. Weckesser, H. López-Carrera, and M. Weidemüller, Phys. Rev. A **94**, 062703 (2016).
[27] J. Perel, R. H. Vernon, and H. L. Daley, Phys. Rev. **138**, A937 (1965).
[28] X. Flechard, H. Nguyen, E. Wells, I. Ben-Itzhak, and B. D. DePaola, Phys. Rev. Lett. **87**, 123203 (2001).
[29] J. Mitroy, M. S. Safronova, and C. W. Clark, J. Phys. B At. Mol. Opt. Phys. **43**, 202001 (2010).
[30] B. M. Smirnov, Physics-Uspekhi **44**, 221 (2001).
[31] M. Cetina, A. T. Grier, and V. Vuletić, Phys. Rev. Lett. **109**, 253201 (2012).
[32] A. Lambrecht, J. Schmidt, P. Weckesser, M. Debatin, L. Karpa, and T. Schaetz, Nat. Photonics **11**, 704 (2017).
[33] K. Ravi, S. Lee, A. Sharma, G. Werth, and S. A. Rangwala, Appl. Phys. B **107**, 971 (2012).
[34] See Supplemental Material.
[35] L. L. Yan, L. Liu, Y. Wu, Y. Z. Qu, J. G. Wang, and R. J. Buenker, Phys. Rev. A **88**, 012709 (2013).